\documentclass[12pt,a4paper]{article}
\usepackage{graphicx}
\usepackage{times}
\usepackage{amsmath}
\title{\bf Global Helioseismology}

\author{Ga\"el Buldgen$^{1,2}$\\
\vspace{0.5cm}\\
\normalsize $^1$ School of Physics and Astronomy, University of Birmingham, Edgbaston, Birmingham B15 2TT, UK.\\ 
\normalsize $^2$ Observatoire de Gen\`eve, Universit\'e de Gen\`eve, 51 Ch. Des Maillettes, CH-1290 Sauverny, Suisse.}
\date{\mbox{}}
\begin{document}
\maketitle
\setcounter{page}{1001}
\pagestyle{plain}
    \makeatletter
    \renewcommand*{\pagenumbering}[1]{%
       \gdef\thepage{\csname @#1\endcsname\c@page}%
    }
    \makeatother
\pagenumbering{arabic}

%
%
\def\bull{\vrule height .9ex width .8ex depth -.1ex}
\makeatletter
\def\ps@plain{\let\@mkboth\gobbletwo
\def\@oddhead{}\def\@oddfoot{\hfil\scriptsize\bull\quad
"How Much do we Trust Stellar Models?", held in Li\`ege (Belgium), 10-12 September 2018 \quad\bull}%
\def\@evenhead{}\let\@evenfoot\@oddfoot}
\makeatother
%
%
\def\beginrefer{\section*{References}%
\begin{quotation}\mbox{}\par}
\def\refer#1\par{{\setlength{\parindent}{-\leftmargin}\indent#1\par}}
\def\endrefer{\end{quotation}}
%
%
{\noindent\small{\bf Abstract:} 
Helioseismology is one of the most successful fields of astrophysics. The observation and character-
ization of solar oscillation has allowed solar seismologists to study the internal structure and dynamics of the
Sun with unprecedented thoroughness. Ground-based networks and dedicated space missions have delivered
data of exquisite quality, enabling the development of sophisticated inference techniques. The achievements
of the fields count, amongst other, the determination of solar photospheric helium abundance, unacessible to
spectroscopic constraints, the precise positioning of the base of the convective zone and the demonstration of
the importance of microscopic diffusion in stellar radiative regions. Helioseismology played also a key role in
validating the framework used to compute solar and stellar models and played an important role in the so-called
solar neutrino problem. In the current era of astrophysics, with the increasing importance of asteroseismology
to precisely characterize stars, the Sun still plays a crucial calibration role, acting as a benchmark for stellar
models. With the revision of the solar abundances and the current discussions related to radiative opacity com-
putations, the role of the Sun as a laboratory of fundamental physics is undisputable. In this brief review, I will
discuss some of the inference techniques developed in the field of helioseismology, dedicated to the exploitation
of the solar global oscillation modes.
}
\vspace{0.5cm}\\
{\noindent\small{\bf Keywords:} Helioseismology -- Solar Physics -- Solar Abundances}
%
%

\section{Introduction}
The study of solar oscillations, coined helioseismology by Douglas Gough, has so far been one of
the most successful field of astrophysics. Thanks to the excellent quality of the data, the precision
of the inferences on the solar structure and the level of details and thoroughness that can be achieved
are unmatched in astronomy. Amongst these successes, one can note the precise determination of the
position of the base of the convective envelope, the determination of present-day helium abundance,
unaccessible to spectroscopy, and the determination of the radial profiles of thermodynamic quantities
as well as the 2D rotation velocity profile inside the Sun \cite{KosovichevRota, BrownRota, Vorontsov91, Dziembowski91, KosovBCZ, JCD91Conv, Antia94, Antia94Sun}.
These inferences have served to validate the framework of the standard solar models \cite{Bahcall82, MODELS, Vinyoles}, which has since been extensively used to model solar-like stars with asteroseismology. Seismic
inference techniques became more and more sophisticated, establishing the reliability of seismology
and its position as a “goldgen path” to study stellar structure and evolution. In the meantime, the
fundamental ingredients of solar and stellar models were being revised, with improvements made on
the equation of state of the solar plasma, the radiative opacities, the transport of chemical elements,... Amongst those revisions, the most problematic one was the reduction of $30\%$ of the solar heavy
element abundance by Asplund et al. \cite{AGSS09} which led to the so-called ``solar modelling problem''.
In this paper, I will give a brief overview of the field of global helioseismology, focussing on the
inferences from normal acoustic modes using seismic inversion techniques. I will not discuss the
field of local helioseismology (see for example \cite{Gizon10} for a review on this topic) which focusses on
studying the properties of the upper convective zone. I will start in section 2 by a brief history of the
observations of solar oscillations and the future of solar missions. I will then discuss the variational
approaches for solar rotation and structure inversions and present some results of these inferences in
sections 3 and 4. Finally, in section 5, I will briefly discuss how innovative inversion techniques and
calibration procedures have been used to gain insights on the solar modelling problem.

\section{Observations}

The first observations of the solar five minutes oscillation were made by Leighton et al. \cite{Leighton62}, us-
ing Dopplergrams from Mt Wilson Observatory and later confirmed by Evans $\&$ Michard \cite{Evans1962}.
Detailed analyses of the oscillations by Frazier \cite{Frazier1968} indicated that the oscillations were not purely
superficial, confirming the analysis of P. Mein \cite{Mein1966}. Theoretical works by Ulrich \cite{Ulrich} and Stein $\&$
Leibacher \cite{SteinLeibacher} suggested that the oscillations could be standing acoustic waves. C. Wolff and Ando
$\&$ Osaki analysed the stability of such oscillations in the observed frequency and wavenumber range
and found that they could be linearly unstable \cite{Wolff1972, Ando1975}. Their nature was however made clear when
F.L. Deubner identified ridges in the wavenumber frequency diagrams \cite{Deubner75}, confirming the fact that
the oscillations were indeed acoustic modes. These observations were independently confirmed by
Rhodes et al. \cite{Rhodes77} who also used them to make inferences on the properties of the solar convective
zone.

Quickly, early works were attempted to constrain the properties of the solar interior using the
observed oscillation frequencies \cite{Scuflaire1975, JCD1976, Iben1976, Rouse1977}. Brookes et al. \cite{Brookes1978} and Servenyi et al. \cite{Severnyi1976}
announced the detection of a long period oscillation ($\approx 160min$) in the solar spectrum which was
found again later in two independent studies. This long-period signal then disappeared from later
observations. The next step was taken by the Claverie et al. \cite{Claverie1979} and by Grec et al. \cite{Grec1980} who provided
observations of modes of low harmonic degrees in the same range of period. Both datasets at low
and high degrees were bridged by Duvall $\&$ Harvey \cite{Duvall1983} who observed low and intermediate degree
oscillation modes. The stage was thus set for the use of solar seismic observations to constrain solar
models (see for example \cite{Scuflaire1975, JCD1976, Gough1986, Brodskii1987, Shibahashi1988} for early works).

In the meantime, observations by Rhodes et al. \cite{Rhodes1979, Deubner1979} allowed to determine for the first time
the rotational splitting of high degree p modes. The analysis was extended by Duvall et al. \cite{Duvall1986} on
intermediate and low degree modes. These observations were then used to infer the properties of the
solar rotation profile (see section 3). With the development of the field, it became clear that long,
uninterrupted time-series where required to provide accurate and precise inferences on the solar inter-
nal structure. The advent of observational programs such as GONG \cite{Harvey1988} and BiSON \cite{Brookes1978,Isaak1989}, using
networks of ground-based observatories provided high-quality data for helioseismic investigations.
With the advent of the SOHO spacecraft \cite{Domingo1995}, in 1996, the quality of the data further improved and
led to the development and extensive use of sophisticated inversion techniques.

\section{Inversions of solar rotation}
\subsection{Formalism}
\begin{figure}[h]
\begin{minipage}{8cm}
\centering
\includegraphics[width=8cm]{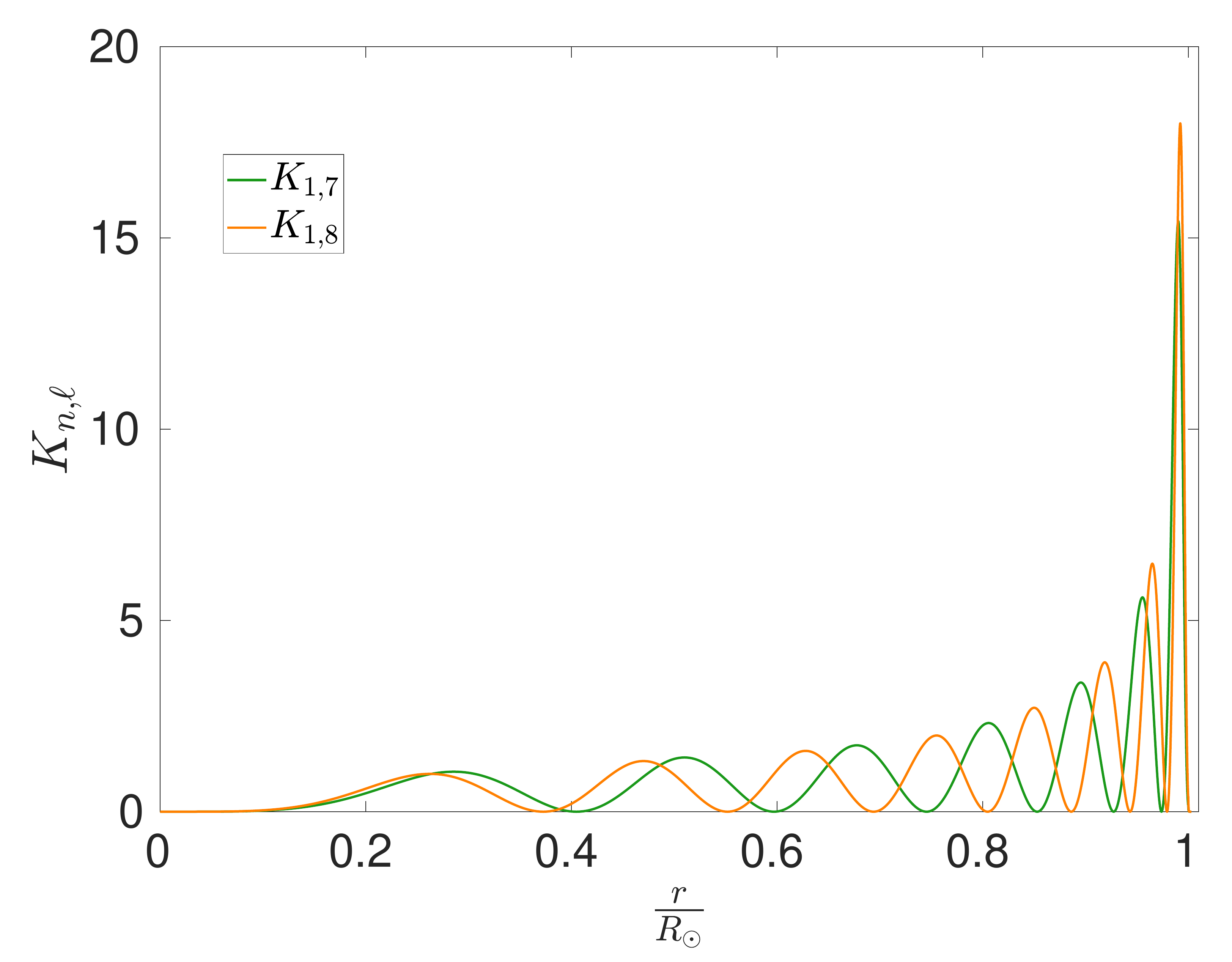}
\caption{1D Rotational kernels for low order
pressure modes of a standard solar model built
using the AGSS09 abundances, FreeEOS equa-
tion of state and OPAL opacities. \label{RotaKerp}}
\end{minipage}
\hfill
\begin{minipage}{8cm}
\centering
\includegraphics[width=8cm]{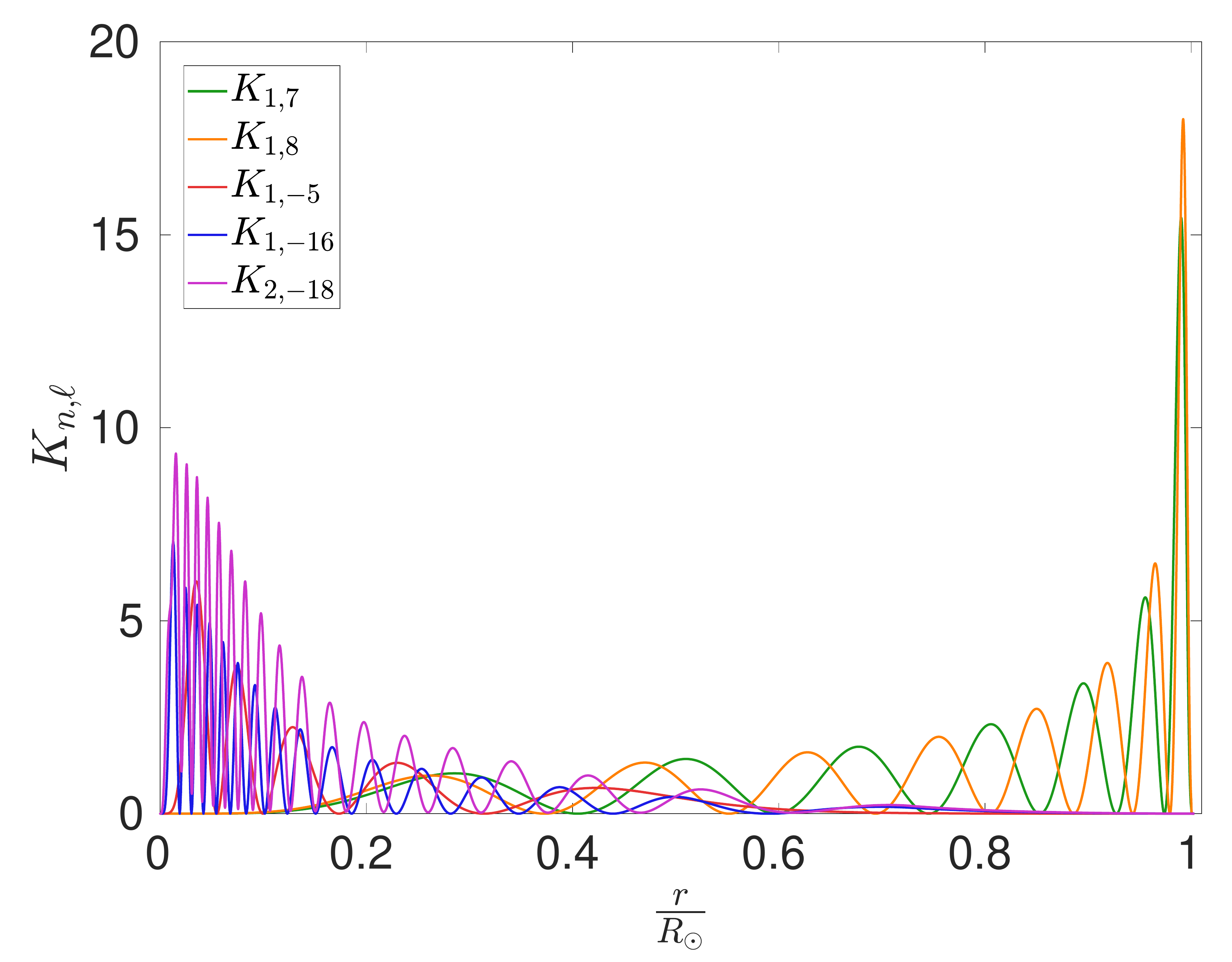}
\caption{Same as Fig. 1, but also including ro-
tational kernels of some gravity modes, showing
their capabilities to strongly constrain the rota-
tion of the solar core. \label{RotaKerg}}
\end{minipage}
\end{figure}

It is well known from the classical theory of stellar pulsations \cite{courscd, Unno} that for a non-rotating star,
the solutions to the eigenvalue problem are degenerate and all the frequencies can be characterised
by two quantum numbers only, their degree, $\ell$ and their radial order, $n$. However, if the rotation of
the star is taken into account, the symmetry of the system will be broken and the solutions have to be
characterised by three quantum numbers, $\ell$, $n$ and the azimuthal order, $m$. In the case of slow rotation,
as for the Sun, these effects can be treated as a perturbation of the spherically symmetric solutions
and to the first order for a rigid rotation. In that case, the changes on the frequencies are written as
\begin{eqnarray}
\nu^{\pm m}_{n,\ell} = \nu^{0}_{n,\ell}\pm m \beta_{n,\ell}\Omega ,
\end{eqnarray}
with $\Omega$ the rotation rate of the star, $\nu^{m}_{n,\ell}$, the frequency including the effect of rotation, $\nu^{0}_{n,\ell}$
of the non-rotating star, and $\beta_{n,\ell}$ the Ledoux constant related to rotation. If one considers a differential
rotation in radius, the rotational splitting will be symmetrical and one can apply a variational analysis
leading to an integral relation between the splitting and the rotational profile. If the horizontal variations of the rotational profile are taken into account, the rotational splitting is not constant anymore.
This integral relation was used in the solar case to carry out inversions of the solar rotation profile. In
the 2D case, the relation reads

\begin{eqnarray}
\delta \omega_{n,\ell,m} = \frac{m \mathcal{R}_{n,\ell}^{m} }{\int_{0}^{R}\rho_{0}\left[\xi_{r,0}^{2}+ \ell(\ell+1)\xi_{h,0}^{2} \right]r^{2}dr},
\end{eqnarray}

with $\mathcal{R}^{m}_{n,\ell}$ defined as follows

\begin{eqnarray}
\mathcal{R}_{n,\ell}^{m} & = \int_{0}^{\pi}\sin \theta d \theta \int_{0}^{R}\left( \vert \xi_{r} \vert^ {2} P_{\ell}^{m}(\cos \theta)^{2} + \vert \xi_{h} \vert^{2}\left[(\frac{d P_{\ell}^{m}}{d \theta})^{2}+ \frac{m^{2}}{\sin^{2} \theta} P_{\ell}^{m}(\cos \theta)^{2} \right] \right.\nonumber \\ 
& \left. - P_{\ell}^{m}(\cos \theta)^{2} \left[ \xi_{r}^{*}\xi_{h} + \xi_{r}\xi_{h}^{*} \right] - 2 P_{\ell}^{m}(\cos \theta) \frac{d P_{\ell}^{m}}{d \theta} \frac{\cos \theta}{\sin \theta} \vert \xi_{h} \vert^{2} \right) \Omega(r,\theta)\rho_{0}(r)r^{2}dr,
\end{eqnarray}

which is often written in the so-called \textit{kernel} form

\begin{eqnarray}
\delta \omega_{n,\ell,m} = m\int_{0}^{R}\int_{0}^{\pi} K_{n,\ell,m}(r,\theta)\Omega (r,\theta)r dr d \theta,
\end{eqnarray}

with $K^{n,\ell,m}$ the rotational kernel associated with the splitting denoted $\delta \omega_{n,\ell,m}$. A review on the various
methods used for rotational inversions can be found in \cite{SchouRota}. Other approaches can be used to express
the rotational splitting, following a decomposition of the splitted frequencies in a series of Legendre
polynomial. We refer the reader to e.g. \cite{Schou1994, Pijpers1997, courscd} and references therein for additional discussions
of such approaches. In figure 1 we illustrate the behaviour of 1D rotational kernel for various p-modes
observed in the solar spectrum and show in Fig \ref{RotaKerg} the important complementary nature of g-modes to
constrain the rotation of the solar core.

The first order approximation is well suited for the Sun which we discuss here, or other typical
slow rotators. However, for the sake of completeness, we mention that the first order variational
analysis is insufficient for faster rotating stars. One solution is to use high order perturbative develop-
ments of the variational approach. It is possible to push further the perturbative developments, lead-
ing to higher order expressions (see e.g. \cite{Dziembowski1992}). It is also possible to develop fully two-dimensional
pulsation codes taking into account the complex geometry of the pulsation modes in fast rotators \cite{Reese2012, Reese2013}
\cite{Reese2017, Reese2017b, Ouazzani2012}. Recently, a non-adiabatic version of such a code has been developed [109, 108].

However, the main issue of the fast rotators is the absence of regularity in their oscillation spectra,
making it almost impossible to identify the modes\footnote{We refer the interested reader to the work by Ligni\`eres et al. \cite{Ligniere2010} and Mirouh et al. \cite{Mirouh2019} and references therein for
further discussion on this topic.}.

\subsection{Results and current questions}

The inversion of the solar rotation profile was first carried out by Brown $\&$ Morrow \cite{BrownRota} and by A.
Kosovichev \cite{KosovichevRota} and was a striking success for helioseismic inferences. Over the years, the improve-
ment of the dataset and of the inversion techniques allowed for a precise 2D cartography of the solar
rotation profile. As can be seen from the results of, for example, Schou et al. \cite{Schou1994, SchouRota}, the rotational
inversions are unable to probe the deep core and the poles. This is essentially linked to the behaviour
of the kernel functions which are poorly localized in both regions. The low intensity of the kernels
in the central regions is a result of the nature of the oscillation, and the same limitations will apply
to the structural inversions. Their inability to probe the polar regions results from their asymptotic
behaviour with lattitude, falling quickly to 0 at the poles. As a result, even a large number of p modes
do not allow to probe efficiently the solar rotation at the poles.
The inversion results of the solar rotational profile were a surprise for stellar modellers, as theoretical prescriptions predicted a steep differential rotation in the solar radiative layers. Consequently, the
solid-body rotation of the solar radiative zone was a surprise and called for investigations of additional
processes for the transport of angular momentum.

A few processes were suggested: fossil fields \cite{Charbonneau1993,Gough1998B,Spada2010}, turbulence-induced gravity waves \cite{Charbonnel2005}
and the dynamo process suggested by H. Spruit \cite{Spruit2002, Eggenberger2005}. Recently, plume-induced gravity waves
have also been suggested by Pincon et al. \cite{Pincon2016,Pincon2017} to reproduce the core rotation of subgiants
observed with Kepler. Hence, as the efficiency of these plume-induced waves has been found to be
higher than the ones generated by turbulence, their impact on the rotation profile of solar models
should also be tested. Currently, all suggested processes can reproduce with more or less the same
degree of agreement the solar rotation profile and thus there is no way to distinguish between them.
There has been some debate as to physical occurence of some of these processes in the litterature but none has been currently ruled out.

As a matter of fact, the solution to this issue is linked to the unraveling of the rotation of the
solar core. One way to achieve this is by detecting solar gravity modes, which would provide such
a diagnostic. In the past, various detections of these modes have been claimed, with the most recent
being made by Fossat et al. \cite{Fossat2017}. However, each detection has been highly debated (see e.g. the work
by Schunker et al. \cite{Schunker2018} and Appourchaux $\&$ Corbard \cite{Appourchaux2019} discussing the Fossat et al. detection) and
it appears that there is still no undisputed detection of the solar gravity modes. We refer to the paper
by Appourchaux et al. \cite{Appourchaux2010} for a review on the history of solar g-mode detections.

Besides classical linear inversions, Corbard et al. \cite{Corbard1999} have also carried non-linear inversions,
to resolve the rotation in the tachocline region. Their approach is based on a generalization of the
“classical” Regularized Least Square inversion technique used in helioseismology, better known as
the Tikohnov regularization. This method uses of a smoothing constraint on the second derivative of
the profile of the inverted quantity, which is inadequate to resolve the sharp variations expected in the
tachocline. In addition, Corbard et al. \cite{Corbard1998} attempted to invert the solar core rotation rate from the
splitting of low degree p modes but found that the results were not robust and thus could not be used
to draw firm conclusions on the solar core rotation rate.

\section{Inversions for solar structure}

As mentioned above, the detection of the solar oscillations paved the way for the development helioseismic inference techniques. The first comparisons were rather simple and consisted in comparing
the frequencies of solar models to the observations. However, it quickly became clear that the small mismatches in frequencies between theoretical models and observations could be used to truly infer
the properties of the solar plasma. The first inversions were based on the asymptotic approximation of stellar adiabatic oscillations and used to infer the sound speed profile in the Sun. We will briefly present this approach in Section 4.1. In Section 4.2, we will discuss the variational formalism of
structural helioseismic inversions and briefly discuss the variables that can be inverted using these equations as well as the limitations of this formalism. 

\subsection{Asymptotic inversions}

The first solar structural inversions were based on the asymptotic expression of Duvall \cite{Duvall1982D}. The
asymptotic analysis of adiabatic oscillations allows to reduce the $4^{th}$ order system to a $2^{nd}$ order
one. This is made by using the so-called Cowling approximation, namely that the perturbation to the
gravitational potential can be neglected in the pulsation equations.

Interesting properties of oscillations in a specific range of $\ell$ and $n$ can be derived in the asymptotic
regime of the stellar oscillations, which drove the first helioseismic inferences and defined helioseis-
mology as an inverse problem. A full description of the mathematical developments related to the
asymptotic regime of adiabatic stellar oscillations can be found in \cite{Shibahashi1979Asymp,Tassoul1980}\footnote{We also refer the reader to the reference textbook by Unno et al. \cite{Unno} and the lecture notes by Jørgen Christensen-
Dalsgaard \cite{courscd} as well as the papers by M. Takata \cite{Takata2016a,Takata2016b} for asymptotic developments applied to mixed modes.}.

Once the Cowling approximation has been used to reduce the order of the system, the equations
can be then rewritten (see \cite{Gough1993Houches} for a more thorough discussion) as a function of a modified displacement function

\begin{eqnarray}
X&=c^{2}\sqrt{\rho} \vec{\nabla}\cdot \vec{\xi}.
\end{eqnarray}

By neglecting the variation of the gravity, $g$ and density, $\rho$ when compared to the perturbed thermodynamical quantities, one can obtain a second order differential equation for X

\begin{eqnarray}
\frac{d^{2}X}{dr^{2}}&=-K(r)X, \label{eq:Asymp2nd}
\end{eqnarray}

with

\begin{eqnarray}
K(r)&=\frac{1}{c^{2}}\left[ S^{2}_{\ell}\left( \frac{N^{2}}{\omega^{2}}-1\right)+\omega^{2} -\omega^{2}_{c}\right],
\end{eqnarray}

where we defined the acoustic cut-off frequency, $\omega_{c}$ as
\begin{eqnarray}
\omega_{c}^{2}&=\frac{c^{2}}{4H_{P}}\left(1-\frac{dH_{P}}{dr} \right).
\end{eqnarray}

If $K(r)$ is positive, the solution will be an oscillating function. However, if $K(r)$ is negative, the solution will show an exponential behaviour. In the surface layers, the dominant term of $K(r)$ will be $(\omega^{2}-\omega^{2}_{c})$ because $S^{2}_{\ell}$ becomes negligible. Consequently, the behaviour of the mode will be dictated by the difference $(\omega^{2}-\omega^{2}_{c})$. If $\omega<\omega_{c}$ the mode will show an exponential decay in the upper regions and thus be trapped in the star. If, in contrast, $\omega$ is larger than $\omega_{c}$, it will show an oscillating behaviour in the atmosphere and thus will lose its energy very quickly.

The analytical solution to equation \ref{eq:Asymp2nd} is found by using the JWKB approximation (standing for
Jeffreys, Wentzel, Kramers, and Brillouin) which was used in quantum mechanics and applied by
Unno et al. \cite{Unno} in the context of stellar pulsations. The fundamental hypothesis is that the solution
will vary faster than the equilibrium quantities. In other words, $X(r)$ varies more rapidly than $K(r)$
and can be described by a function of the form
\begin{eqnarray}
X(r)&=a(r)e^{i\phi(r)},
\end{eqnarray}

where $\phi(r)$ varies much faster than $a(r)$ and one can derive a local wavelength of the form

\begin{eqnarray}
n&=\frac{d\phi}{dr}.
\end{eqnarray}

This solution can be inserted in equation \ref{eq:Asymp2nd} and one can find that the behaviour of the solution will
again be sinusoidal or exponential depending on the sign of $K(r)$. After some additional mathematical
developments and the proper treatment of the boundary conditions at the reflexion points, one can
derive the asymptotic form of the eigenfunctions and show that the frequencies of modes trapped
between two turning points $r_{1}$ and $r_{2}$ must satisfy the following relation

\begin{eqnarray}
\int_{r_{1}}^{r_{2}}K^{1/2}(r)dr &= \left(n-\frac{1}{2} \right)\pi. \label{eqDispAsymp}
\end{eqnarray}

For pressure modes which have $\omega^{2}\gg\vert N \vert^{2}$, it can be shown that equation \ref{eqDispAsymp} reduces to the so-called Duvall law [57]
\begin{eqnarray}
\int_{r_{t}}^{R} \left( 1-\frac{(\ell (\ell +1))^{2}c^{2}}{\omega^{2}r^{2}}\right) \frac{dr}{c} &= \frac{(n+\alpha(\omega))\pi}{\omega}, \label{eqDuvallAs}
\end{eqnarray}
with $\alpha(\omega)$ depending on the surface regions, $r_{t}$ being the lower turning point where $S_{\ell}(r_{t})=\omega$ and $R$ the upper turning point where $\omega_{c}(R)=\omega$, which is valid for moderate values of $\ell$. 

This expression can then be used to infer the sound speed profile of a solar model. Equation
12 is an integral equation of the Abell type, which can be inverted analytically to infer the sound
speed profile. A generalization of this expression for high $\ell$ modes has been derived by Vorontsov
$\&$ Shibahashi \cite{Vorontsov1991}. In practice, various techniques have been developed to invert the Duvall law
\cite{Gough1984, JCD1985, Gough1986, Brodskii1987, Brodsky1988, Shibahashi1988}, with some using a differential formulation of the dispersion relation \cite{JCD1989DiffAS, JCD91Conv}.
\subsection{Variational approach}

The most commonly used formalism to carry out inversions of the solar structure is based on the
variational principle of stellar adiabatic oscillations, which has been derived in a very general context
by \cite{Chandrasekhar1964, LyndenBell}. The variational formalism can be used to describe the relation between frequency pertur-
bations and structural corrections in relatively simple form. This was shown by Dziembowski et al.
\cite{Dziemboswki90}, who wrote the now classical integral relations between frequency perturbations and sound speed
and density corrections:

\begin{eqnarray}
& \frac{\delta \omega_{n,\ell}}{\omega_{n,\ell}} = \int_{0}^{R}\left[K^{n,\ell}_{c^{2}, \rho} \frac{\delta c^{2}}{c^{2}} + K^{n,\ell}_{\rho,c^{2}} \frac{\delta \rho}{\rho}\right]dr,\label{eq:Varrhoc}
\end{eqnarray}
where we have defined the $K^{n,\ell}_{c^{2}, \rho}$, $K^{n,\ell}_{\rho,c^{2}}$, the structural kernels of the $(\rho,c^{2})$ structural pair. It should be noted that these functions are only depending on unperturbed variables, thus only on the theoretical model that is built to carry out the inversion. The mathematical expression of the structural kernels is the following
\begin{eqnarray}
& K^{n,\ell}_{c^{2}, \rho}  = \frac{c^{2}\rho \Lambda^{2}_{\ell}(r)r^{2}}{2E_{n,\ell}\omega_{n,\ell}^{2}}, \label{eq:A33} \\
& K^{n,\ell}_{\rho,c^{2}}  = \frac{1}{2E_{n,\ell} \omega_{n,\ell}^{2}}\left( -\omega^{2} \rho \left[ \xi_{r}(r)^{2}+\ell(\ell+1) \xi_{h}(r)^{2}\right] r^{2} + 2 \rho (\xi_{r}\frac{d \phi^{'}}{dr} + \frac{\ell(\ell+1)}{r}\xi_{h}\phi^{'}) r^{2}\right. \nonumber \\
 & - 2 \xi_{r} \Lambda_{\ell}  \rho G m(r)+ 4 \pi G \rho^{2}\xi_{r}^{2} r^{2} + 2 \xi_{r} \frac{d \xi_{r}}{dr} G  \rho m(r) \nonumber \\ & \left.- 4\pi \rho G
(\int_{r}^{R} \xi_{r}^{2}\frac{d \rho}{dr} + 2 \Lambda_{\ell} \rho  \xi_{r} d\tilde{r})r^{2} + \Lambda_{\ell}^{2} c^{2}\rho r^{2} \right), \label{eq:A34}
\end{eqnarray}
where 
\begin{eqnarray}
E_{n,\ell} = 4\pi \int_{0}^{R}\left[ \vert \tilde{\xi}^{n,\ell}_{r}(r)\vert^{2}+\ell(\ell+1)\vert \tilde{\xi}^{n,\ell}_{h}(r)\vert^{2}\right]\rho r^{2}dr. \label{eq:A35}
\end{eqnarray}

This relation has since been extensively exploited to test the internal structure of standard solar models, confirming their good agreement and thus the relative accuracy of our depiction of solar structure.

\begin{figure}[h]
\centering
\includegraphics[width=16cm]{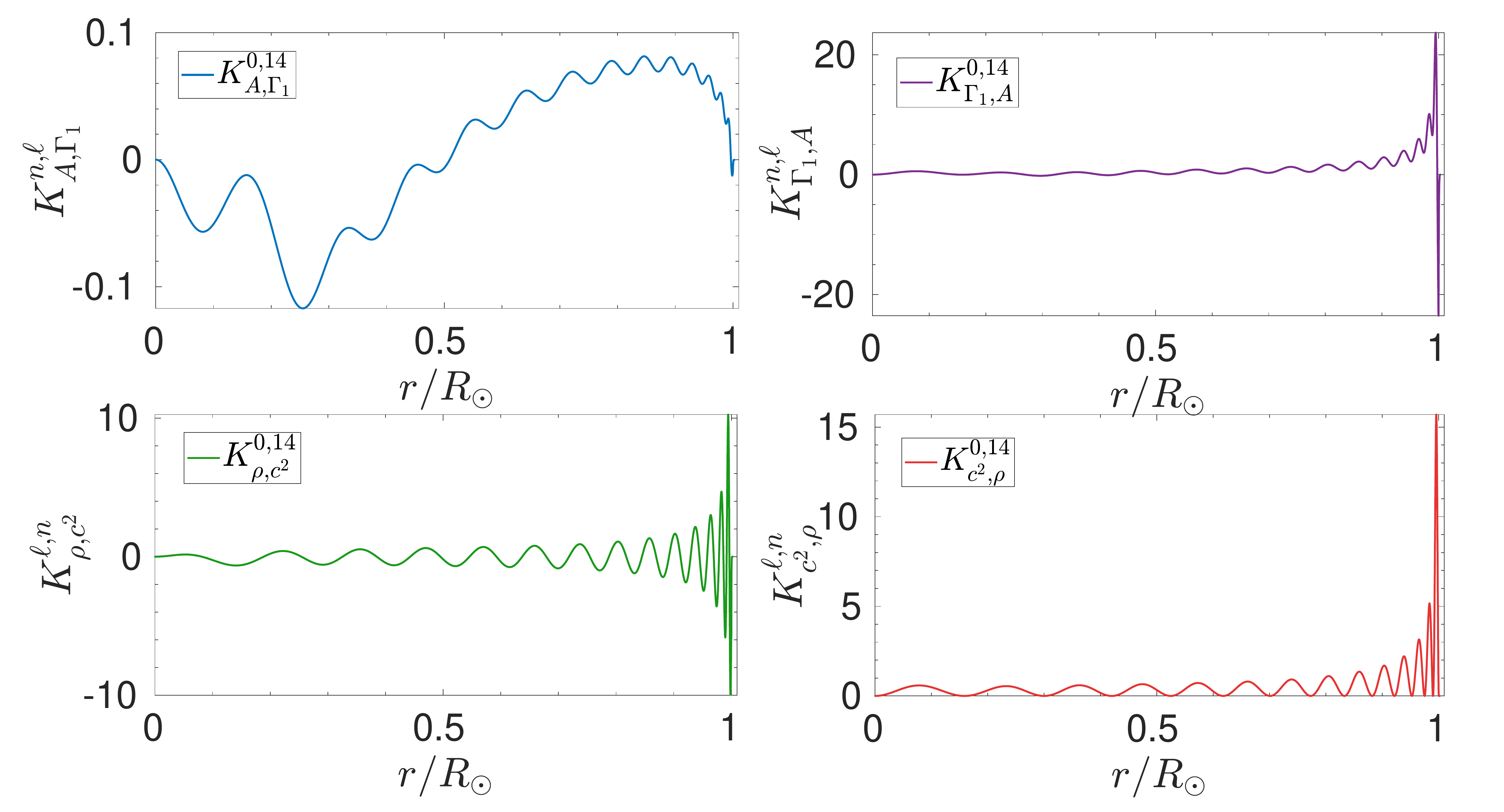}
\caption{Example of structural kernels for the $(A,\Gamma_{1} )$ structural pair (upper figures) and for the $(c^{2},\rho)$
structural pair (lower figures). \label{figStrucKernels}}
\end{figure}

The variational equations can be modified, implying that one has access to more than the just the
sound speed and the density profile of the Sun. This can be done using simple algebra to obtain the
kernels related to density and $\Gamma_{1}=\frac{\ln P}{\partial \ln \rho} \vert_{S}$ perturbations. Other mathematical ``tricks'' can be used
to derive kernels for less obvious structural quantities, such as the Ledoux discriminant or an entropy
proxy, defined as $S_{5/3}=\frac{P}{\rho^{5/3}}$
\cite{BuldgenS,BuldgenA}. These methods have been recently used to perform extensive tests of the internal structure of the Sun, in light of the revision of the solar metallicity by Asplund et al. \cite{AGSS09}.
We illustrate some of the structural kernels used for the inversion of the solar structure in figure \ref{figStrucKernels}.

\begin{figure}[h]
\centering
\includegraphics[width=13cm]{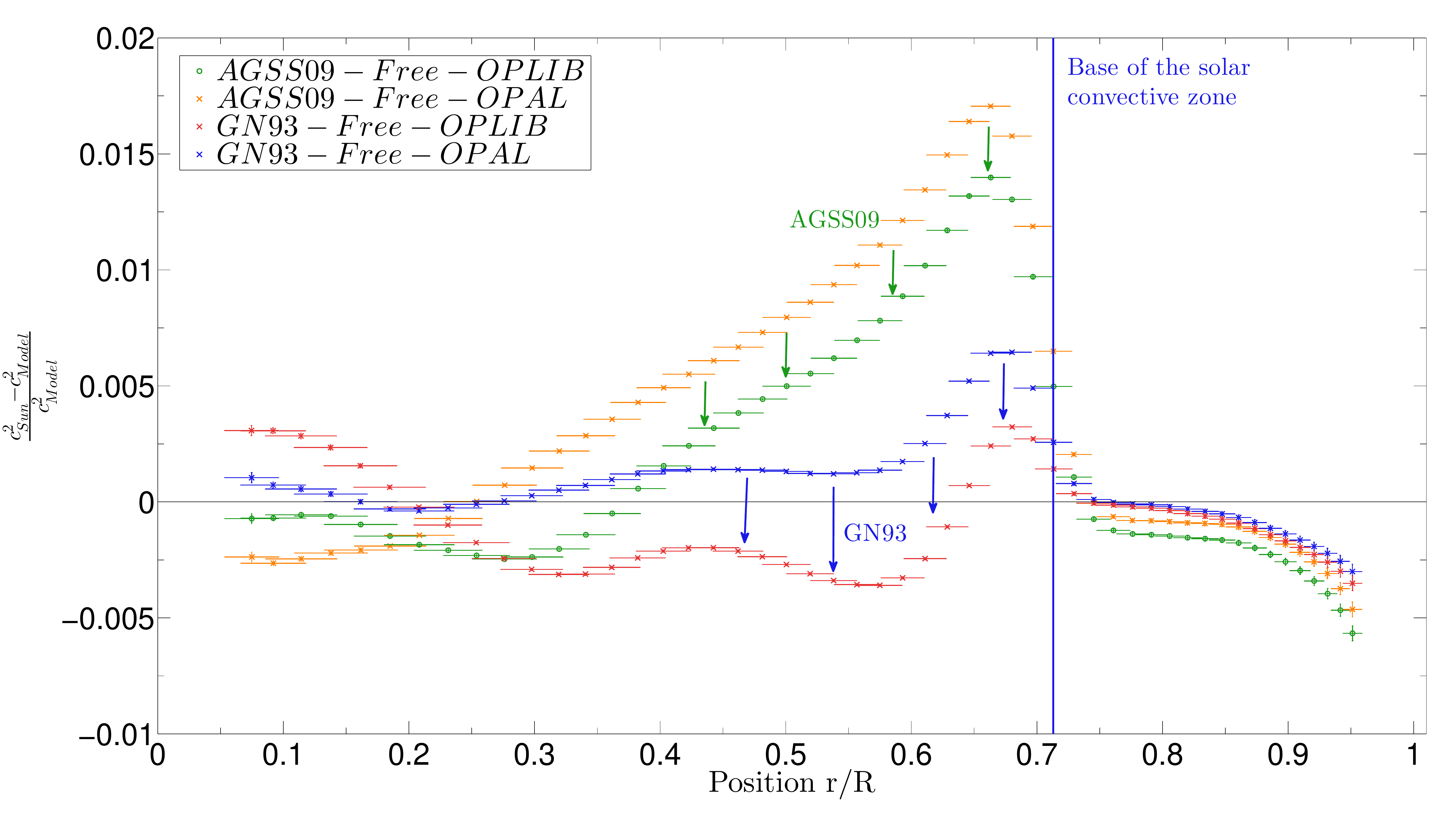}
\caption{Squared adiabatic sound speed inversions for standard solar models built using the AGSS09
and GN93 abundances combined with both the OPAL and OPLIB opacity tables. \label{figc2InvBuldgen}}
\end{figure}

\begin{figure}[h]
\centering
\includegraphics[width=13cm]{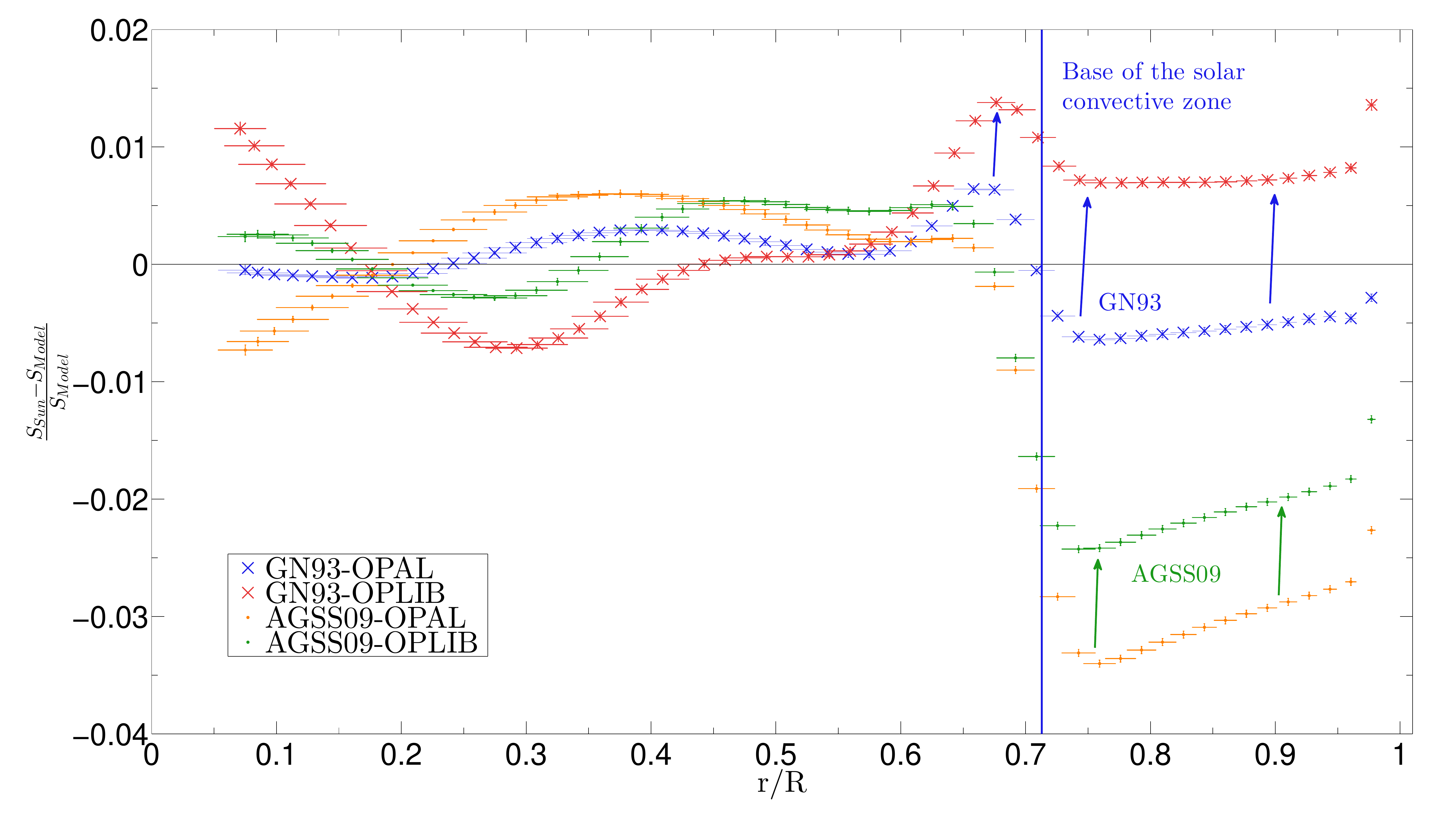}
\caption{Entropy proxy inversions for standard solar models built using the AGSS09 and GN93
abundances combined with both the OPAL and OPLIB opacity tables. \label{figSInvBuldgen}}
\end{figure}

\begin{figure}[h]
\centering
\includegraphics[width=13cm]{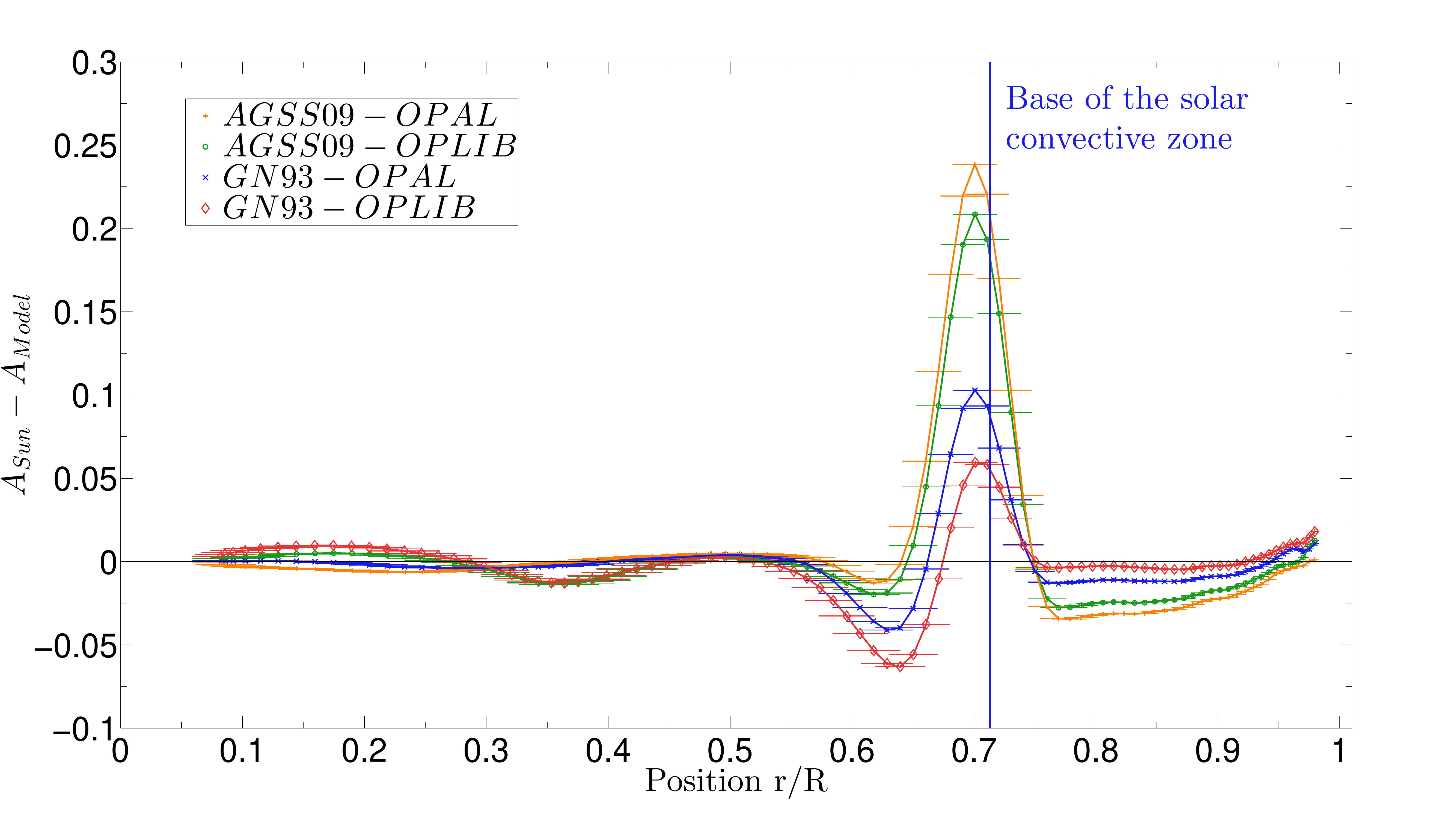}
\caption{Ledoux discriminant inversions for standard solar models built using the AGSS09 and GN93
abundances combined with both the OPAL and OPLIB opacity tables. \label{figAInvBuldgen}}
\end{figure}

As an illustration, we also show in figures \ref{figc2InvBuldgen}, \ref{figSInvBuldgen} and \ref{figAInvBuldgen} some inversion results for solar models built
using both the AGSS09 and GN93 abundances combined with both the OPAL and OPLIB opacities,
illustrating the changes that resulted from the revision of the solar opacities. Similar results can be
found in LePennec et al. \cite{LePennec} for the OPAS opacities and a more extensive investigation of various
inversion results can be found in Buldgen et al. \cite{Buldgen2019Sun}. The dataset used for these inversions is a
combination of MDI and BiSON data \cite{BasuSun,Davies}.

Thanks to these generalized formalisms, any variable that can appear in the adiabatic pulsation
equations can be probed using seismic inversions. This enables to focus the seismic information on
various aspects of the solar structure. The variational approach can also be extended to derive so-
called secondary variables, which do not directly appear in the equations of adiabatic oscillations
\cite{Gough1988Stanford}. Usually, this is done by assuming that the equation of state of the solar plasma is known.
Using this approach, one can either directly develop the perturbations of the ``acoustic'' variables
into perturbations of the secondary thermodynamic variables that one wishes to infer, or to fit the inverted results of the “acoustic” variables in the computation of a static solar model (see for example
\cite{Takata1998,Takata2003,Gough2006,Shibahashi2006}). The variational developments of quantities related to the equation of state have
also been used to seismically infer the helium abundance in the convective zone, using perturbations
to the adiabatic exponent $\Gamma_{1}$ following the formula

\begin{eqnarray}
\frac{\delta \Gamma_{1}}{\Gamma_{1}}=\frac{\partial \ln \Gamma_{1}}{\partial \ln P}\vert_{\rho,Y,Z}\frac{\delta P}{P} + \frac{\partial \ln \Gamma_{1}}{\partial \ln \rho}\vert_{P,Y,Z}\frac{\delta \rho}{\rho} + \frac{\partial \ln \Gamma_{1}}{\partial Y}\vert_{P,\rho,Z} \delta Y + \frac{\partial \ln \Gamma_{1}}{\partial Z}\vert_{P,\rho,Y} \delta Z. \label{eq:Gamma1EosRap}
\end{eqnarray}

with $P$ the pressure, $\rho$ the density, $Y$ the helium and $Z$ the heavy-elements mass fractions. The
main weakness of equation 17 is its strong dependency on the equation of state of the solar plasma.
This was already noted in early investigations \cite{Dziembowski91,RichardY}, who noted that this was the largest source
of uncertainties. Even the recent investigations by Vorontsov and collaborators \cite{Vorontsov13} show relatively
large uncertainties in their inferences of the chemical composition of the solar envelope, despite the
excellent quality of the seismic data and the thoroughness of the investigation. To circumvent the
problem, other authors have used the adimensional sound speed gradient, defined as
\begin{eqnarray}
W(r)=\frac{1}{g}\frac{dc}{dr},
\end{eqnarray}
to constrain the helium abundance in the solar envelope, with $g = \frac{Gm}{r^{2}}$. While this quantity is certainly less directly sensitive to the equation of state, the inference still relies on the reproduction of the $\Gamma_{1}$ properties in the helium ionization regions, which are undoubtedly influenced by the properties
of the solar equation of state.
\begin{figure}[h]
\centering
\includegraphics[width=11cm]{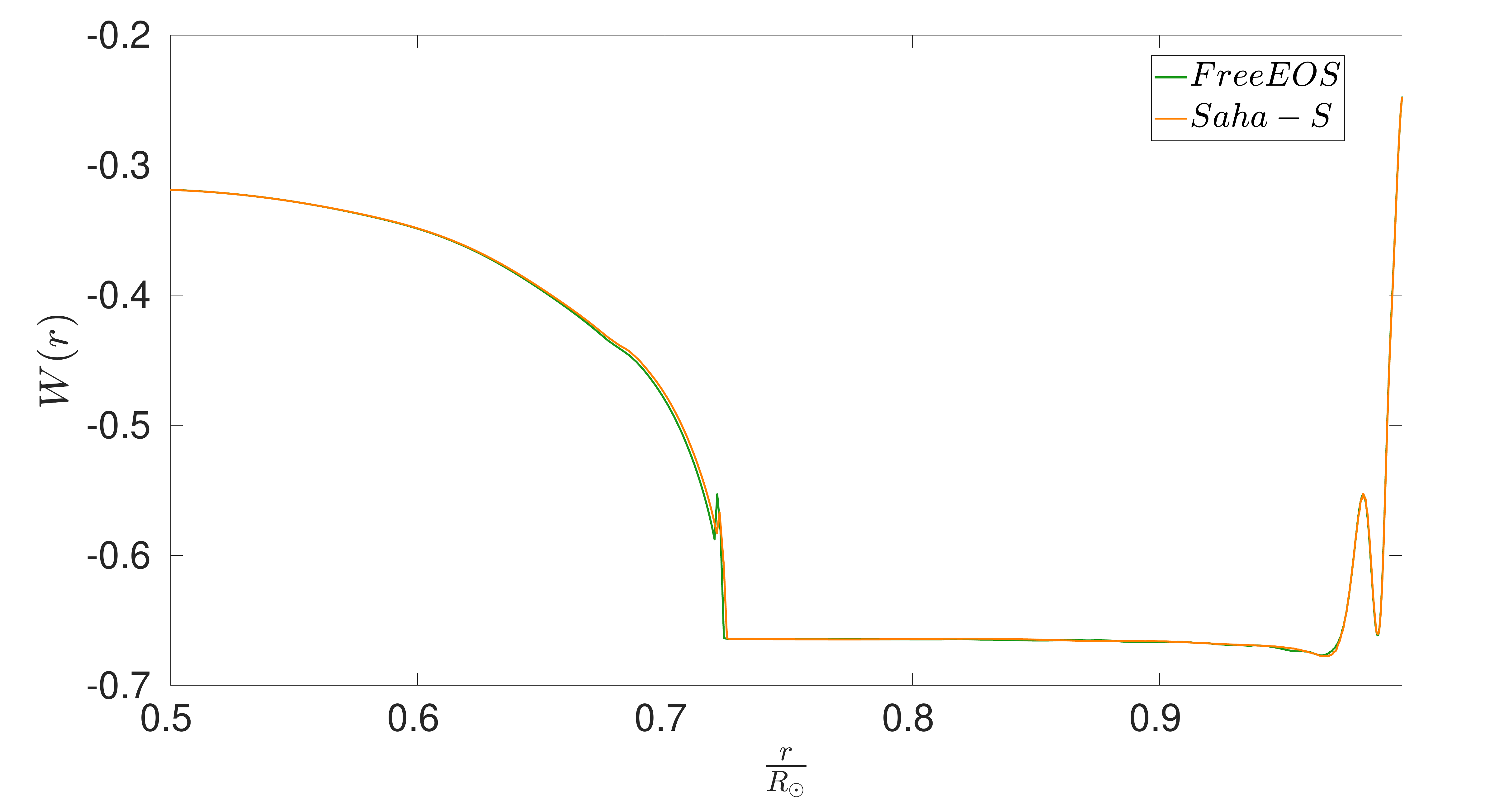}
\caption{Adimensional sound speed gradient as defined by equation 17 for standard solar models built
using the AGSS09 abundances, the OPAL opacities and both the FreeEOS and SAHA-S equation of
state. \label{figWr}}
\end{figure}

Seismic inversions have also been used to infer constraints on the heavy element abundances in
the solar envelope. The first of such an investigation was presented in Takata $\&$ Shibahashi \cite{Takata2001},
before the revision of the solar abundances. Elliott also investigated the behaviour of $\Gamma_{1}$ to determine the agreement of various equations of state to helioseismic results [64]. They report a good agreement
for the GS98 metallicity value but do not attempt to infer it directly from the inverted $\Gamma_{1}$ profile, given its
strong dependency in the equation of state. Since then, various groups have made similar attempts,
following different approaches. Basu $\&$ Antia attempted to determine the heavy elements abundance in the
solar convective region by analysing the behaviour of $W(r)$ and found an abundance in agreement with the GS98 determination \cite{Basu2004}.

More recently, Vorontsov et al. \cite{Vorontsov13,VorontsovSolarEnv2014} pointed out the limitations of the study by Antia $\&$ Basu
and Basu $\&$ Antia \cite{Basu2004,Antia2006}, stating that the equation of state they used was not suitable for such detailed
investigations. They carried out an extensive study of the properties of the solar convective envelope
and determined a low value for the solar metallicity, in agreement with the Asplund value \cite{AGS04O, AGSS09}. In
2017, Buldgen et al. \cite{BuldgenZ} carried out an independent study and confirmed the results of Vorontsov et
al. \cite{Vorontsov13}, noting however the strong dependency on the equation of state and on the properties of the underlying reference model.
\begin{figure}[h]
\centering
\includegraphics[width=13cm]{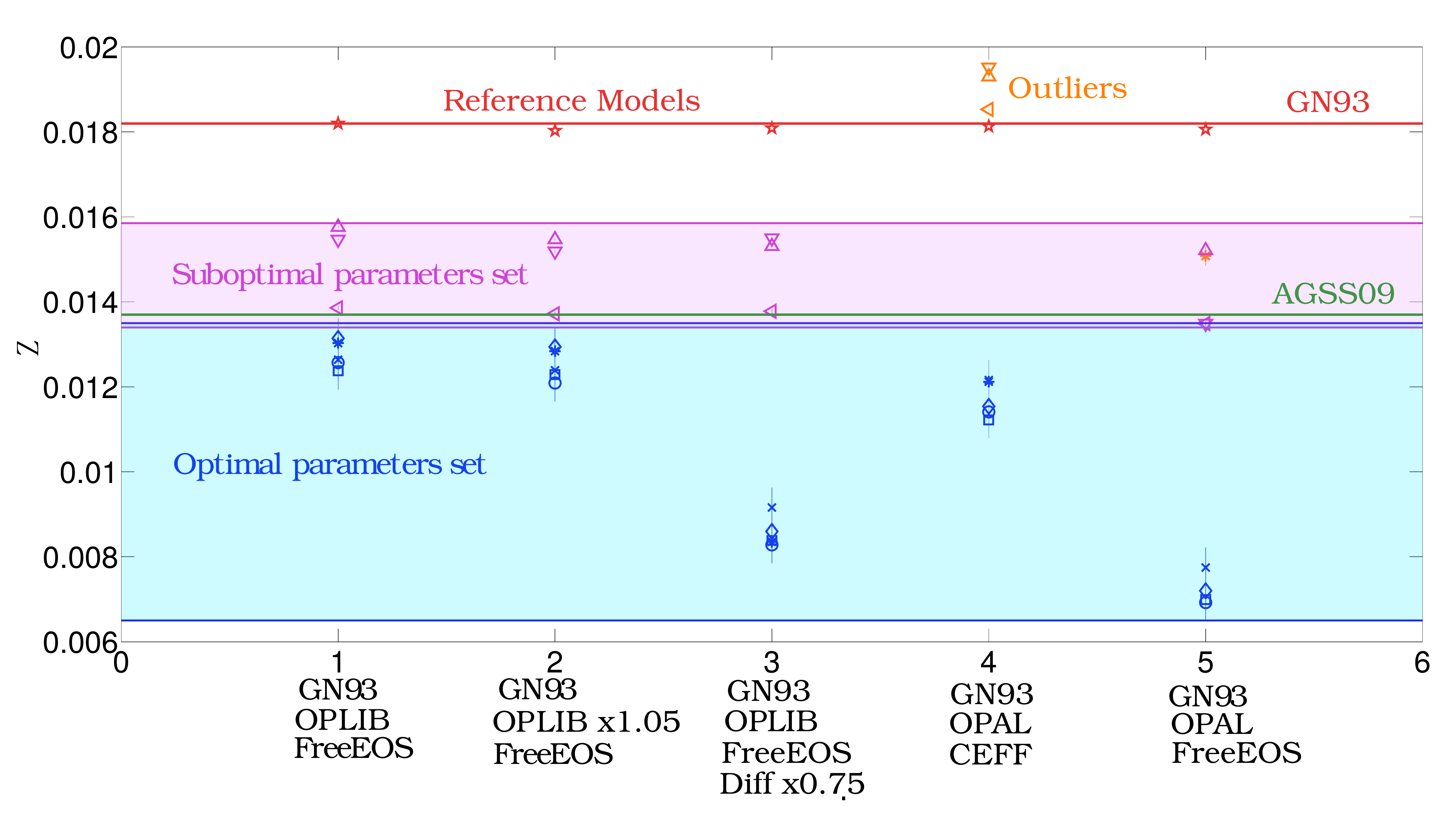}
\caption{Results of metallicity inversions for various calibrated solar models. The inversion parameters have been changed to analyse the stability of the inversion results (See \cite{BuldgenZ} for more details). \label{figZInvBuldgen}}
\end{figure}

The results of seismic inversions have been used to build so-called seismic solar models \cite{Basu1996,Takata1998,Gough2006}. These models do not stem from a standard calibration using evolutionary sequences, but from a
reconstruction of the solar structure using helioseismic data assuming hydrostatic and thermal equi-
librium. These models have been used to infer the properties of the solar tachocline \cite{Takata2003}, constrain
solar axions \cite{Watanabe2002,Shibahashi2003Axions}, but also as ``helioseismic'' references for comparisons to solar models of various groups.

Given the importance of opacity to infer properties of the solar structure, the variational formal-
ism has also been extended to the computation of so-called ``opacity kernels'' \cite{JCD98opacK}, which are use to
express the discrepancies of acoustic variables in terms of opacity modifications. These are imple-
mented following a perturbative approach, considering a linear response of the structural properties
of a model to opacity changes

\begin{eqnarray}
\frac{\delta c}{c}=\int K_{c}(r,T) \delta \log \kappa (T) d \log T, \\
\delta \ln \phi_{\nu} = \int dK_{\phi}(r,T) \delta \log \kappa (T) d \log T. 
\end{eqnarray}

Examples of these opacity kernels can be found in \cite{Villante2010,Vinyoles}. The main hypothesis of this approach
is that the models will respond linearly to the opacity changes, which is valid within a small range of opacity modifications. Moreover, such variations are static inferences and may not reflect the impact of opacity changes over the whole duration of a solar evolutionary sequence.

Analysing these opacity changes is of particular importance in the context of the solar modelling problem and the potential sources of uncertainties in the Sun that may contribute to the current discrepancies. We will further discuss these points in section 5.

One major difficulty when computing any structural helioseismic inversion is dealing with the
so-called surface effects. The issue is well-known in both helio- and asteroseismology and results
from the shortcomings of both the computation of solar and stellar oscillations, and that of the stellar
models. It is now well-known that the solar pressure modes are generated in the upper parts of the
turbulent convective envelope. Hence, the understanding of the driving and damping of the oscillations is intrisically linked to the turbulent closure problem the upper layers of the star, as well as to the
non-adiabatic properties of convection, both of which are poorly modelled and understood in stellar
conditions.

Thus, the surface effect is often said to arise from both a model and a modal contribution, the
former being related to the poor modeling of the properties of the equilibrium structure of the upper
parts of convective envelopes and the latter being related to the non-adiabatic properties of the oscillation modes, resulting from the driving and damping by turbulent convection. In nearly all studies of
solar-like oscillations, including the Sun, oscillations are computed using the hypothesis of adiabaticity, which is justified in the deep layers of the star only. The variational formalism is no exception,
as the hypothesis of adiabaticy is required to derive the symmetry properties of the operator of stellar
oscillations, leading in turn to the integral expressions used in the inversions. Furthermore, a few
boundary terms of integration by parts are neglected in the derivation of the integral relations between
frequency and structure, further contributing to inaccuracies in the surface layers.

The surface effects have been thoroughly studied in the solar case and are most often empirically
modelled following a parametrization of the observed trend. From a theoretical point, one can im-
pose that the perturbation generated from the surface effects, that we will note $\delta D_{surf}$ , is only of high amplitude in the surface layers. At a given frequency, it is possible to show that the behaviour of the
eigenfunctions does not depend on $\ell$ in the surface regions, especially for low degree modes.

As a first approximation, it is assumed that the eigenfunctions do not depend on $\ell$ and this allows
to define an equation similar to equation 13 for the surface effect
\begin{eqnarray}
\frac{< \vec{\xi}_{n,\ell}, \delta \mathcal{D}_{surf}(\vec{\xi}_{n,\ell}) >}{2 \omega_{n,\ell}^{2}}= E_{n,\ell} \frac{\delta \omega_{n,\ell}}{\omega_{n,\ell}}. \label{eq:VariaSURF}
\end{eqnarray}

This equation is formally very similar to equation 13, since we are trying to determine the impact of
a perturbation of the model on the frequencies. In this particular case, the perturbation is located in
the surface regions and equation 20 will thus behave like the eigenfunctions in these regions. In other
words, it will not be strongly dependent on $\ell$, especially if the modes are of low degree. We follow
here the developments of Christensen-Dalsgaard \cite{courscd}, introducing a $Q_{n,\ell}$ function defined as

\begin{eqnarray}
Q_{n,\ell} = \frac{E_{n,\ell}}{\bar{E}_{0}(\omega_{n,\ell})},
\end{eqnarray}

with $\bar{E}_{0}$ being the inertia of the radial mode at a fixed $\omega$, interpolated in $\omega_{n,\ell}$. One obtains the following
relations

\begin{eqnarray}
Q_{n,\ell} \delta \omega_{n,\ell} = \frac{< \vec{\xi}_{n,\ell}, \delta \mathcal{D}_{surf}(\vec{\xi}_{n,\ell}) >}{2 \omega_{n,\ell}\bar{E}_{0}(\omega_{n,\ell})}.
\end{eqnarray}

Christensen-Dalsgaard \cite{courscd} illustrates the effects of this scaling for a comparison between a reference
model and a model with a modified opacity in the upper regions. Consequently, this implies that
the quantity $Q_{n,\ell} \delta \omega_{n,\ell}$ is largely independent of $\ell$ and, in turn, the surface corrections will also only
depend on $\omega$ for a given $\ell$. A first correction is thus to apply this scaling method to eliminate the
$\ell$ dependency of the surface effects. However, this scaling is not sufficient to correct the observed
biases.

From the analysis of the variational equation 13, we know that no particular care is taken to
account for the surface effects. Indeed, the hypotheses of equation 13 are not satisfied in the surface
regions and these inaccuracies mean that there is no theoretical expression for $D_{surf}$. Thus, we have
to add another function to the integral relations, attempting to take into account surface effects. This
function is usually denoted $\mathcal{F}(\omega)$. Using the $Q_{n,\ell}$ factor to normalise the expression, one gets a surface
term independent of $\ell$. In fact, $\mathcal{F}(\omega)$ is simply an empirical modelling of the effects of the operator
$\delta D_{surf}$ , whose form is unknown. For the $(\rho, c_{2})$ structural pair, one gets the following relation

\begin{eqnarray}
& \frac{\delta \omega_{n,\ell}}{\omega_{n,\ell}} = \int_{0}^{R}\left[K^{n,\ell}_{c^{2}, \rho} \frac{\delta c^{2}}{c^{2}} + K^{n,\ell}_{\rho,c^{2}} \frac{\delta \rho}{\rho}\right]dr + \frac{\mathcal{F}(\omega)}{Q_{n,\ell}}. \label{eq:surf}
\end{eqnarray}

The inversion technique now includes an additional function, taking into account the surface effects
that need to be either modelled or eliminated. In practice, it can also be shown that $\mathcal{F}$ must be a slowly
varying function of frequency because any sharp variation in the structure leads to an oscillating
signature in the frequencies whose frequency is proportional to the acoustic depth of the perturbation.
In the case of surface effects, this signature is going to be very slowly varying.

Usually, the $\mathcal{F}$ function is modelled using Legendre polynomials to reproduce the surface effect
and including in the inversion technique an additional condition imposing

\begin{eqnarray}
\sum_{i}c_{i}(r_{0})\mathcal{F}(\omega_{i}) = 0.
\end{eqnarray}

This additional condition thus states that the model of the surface effect should be simultaneously
cancelled by the inversion as it computes the inversion coefficients used to recombine the frequencies.
In practice, the series of Legendre polynomial goes up to order 6 or 7. One should note that there is
no physical justification behind the choice of the Legendre polynomials and that one rather speaks of
``well chosen functions\footnote{See \cite{courscd} for some additional discussion on this topic.}''.

Another method to model surface effects is to use a low-pass filter on the oscillation data \cite{Basu1996Filter}.
One then uses an asymptotic form of the relation between frequency and structure, in the form

\begin{eqnarray}
S_{i}\frac{\delta \omega_{i}}{\omega_{i}}\simeq \mathcal{H}_{1}(\frac{\omega_{i}}{L})+ \mathcal{H}_{2}(\omega_{i}),\label{eq:surf1}
\end{eqnarray}
where $L=\ell+1/2$. This expression is derived from the perturbative analysis of the asymptotic relations of pressure modes given in equation \ref{eqDuvallAs}. The exact analytical expression of the functions $\mathcal{H}_{1}$ and $\mathcal{H}_{2}$ can be found in \cite{JCD1988H2}. The filtering is done in three steps, with the hopes of delivering a correction for the surface effects in the form
\begin{eqnarray}
\sum_{i}a_{i}S_{i}\frac{\delta \omega_{i}}{\omega_{i}}=\mathcal{F}(\omega_{i}). \label{eq:surf2}
\end{eqnarray}
First, one fits a spline combination to equation 25. This allows to obtain an equation linking frequency corrections to $\mathcal{H}_{2}(\omega)$.

The second step is then to apply a low pass filter to this relation to isolate any slowly varying
function of $\omega$. One then obtains a filtered function, $\mathcal{H}_{2}(\omega)$, that is supposed to be linked to $\mathcal{F}(\omega)$.

Thirdly, one fits this filtered $\mathcal{H}_{2}(\omega)$ to obtain a relation similar to equation 26. This fit then allows
to define coefficients to correct the individual frequencies such that one eliminates the surface effects
based on the asymptotic fit. One should note that due to this correction, the structural kernels are
modified and have much lower amplitudes in the surface regions. This method has the disadvantage of
introducing correlations between frequency differences, meaning that the treatment of the propagation
of errors is also slightly more expensive numerically.

\section{The solar modelling problem}

Following the revision of the abundances by \cite{AGS04O,AGS05C, Asplund05}, the helioseismic community was confronted to
a significant reduction of the agreement of standard solar models with inverted profiles. The revised abundances were subject to some controversy and various claims stated that they disagreed with
helioseismology. Further work by Asplund and collaborators \cite{AGSS09} confirmed the revision and thus the
``solar metallicity problem''. Caffau and collaborators \cite{Caffau}, using different hydrodynamical model than the group of M. Asplund, carried out a re-analysis of solar spectra and found an intermediate
value for the solar metallicity. Further investigations were made to determine whether the origin of
the discrepancies stemmed from the atmosphere models. These analyses concluded that the small
differences in the models could not explain the abundance differences and that the discrepancies
between abundance determinations were likely due to the use of different spectral lines, which, in the
case of blends \cite{Allende2001}, could cause an overestimation of the chemical abundances.

The solar problem has since been a tedious issue, as the solution might stem from various con-
tributions of constituents of the standard solar model such as the opacities, the equation of state, the
computation of microsopic diffusion and the heavy element abundance itself, since it is of course
subject to uncertainties. However, the issue has also led modellers to discuss extensively the lim-
itations of the standard model framework. For example, the potential of overshooting at the base
of the convective zone to reconcile low metallicity models with helioseismic constraint (see e.g. \cite{Serenelli2004, Bahcall2005, Guzik2005, Guzik2006, Montalban2006, Delahaye2006, SerenelliComp}), or scenarios involving accretion of material or intense mass loss in the early days of the solar evolution (see e.g. \cite{Castro2007, Wood2018}).

To this days, no clear solution to the solar modelling problem has emerged, despite the publication of updated opacity tables by various groups \cite{Mondet,Colgan}. In 2015, Bailey and collaborators \cite{Bailey} published results of the first experimental measurements of iron opacity in the physical conditions of the base of the convective envelope. These results led to intense discussions in the opacity community (see
e.g. \cite{Nahar, Blancard2016, NaharReply, Krief2016, Iglesias2015,Pradhan}), as the measurements suggested a very significant underestimation of the opacity by theoretical calculations. While theoretical calculations are ongoing \cite{Pain2019, Zhao, Krief2018}, it is
also important to point out the efforts of independent experimental determinations \cite{Ross2016, Heeter2017, PERRY2017, Cardenas2018, Colaitis2018, Dodd2018}.

Besides the opacity issue, a recent revision of the neon over oxygen ratio from analyses of the
solar corona has also been concluded by two independent studies \cite{Landi,Young}. This revision significantly
improves the agreement of AGSS09 models with helioseismic constraints, although not back to the level of agreement of GS98 models.

\begin{figure}[h]
\centering
\includegraphics[width=13cm]{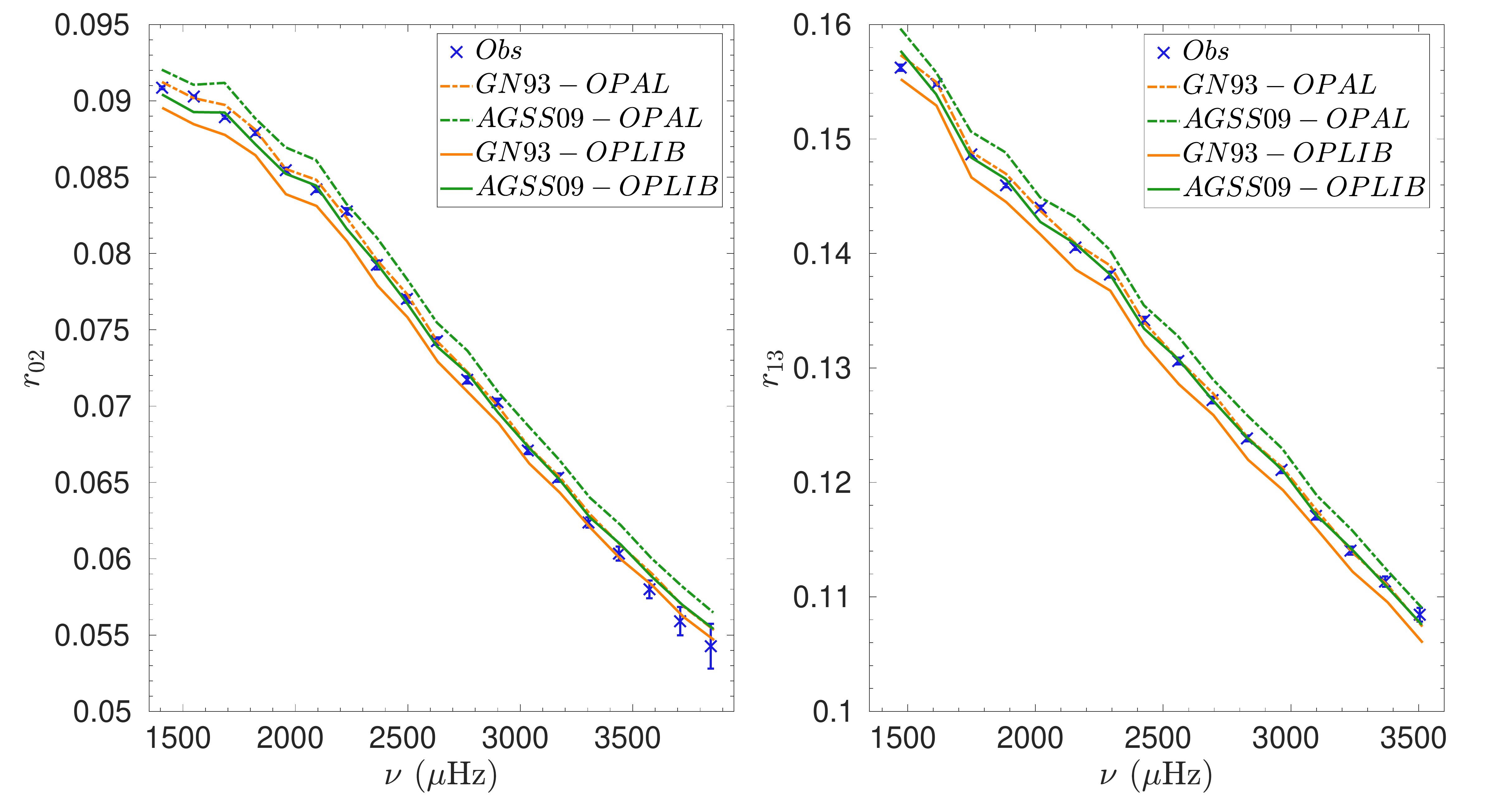}
\caption{Frequency ratios of low order pressure modes observed by the BIrmingham Solar Oscillation Network compared to standard solar models built using both the GN93 and AGSS09 abundances
combined with the OPAL and OPLIB opacities. \label{figRatios}}
\end{figure}

The main difficulty in trying to constrain the solar modelling problem using helioseismology
stems from the fact that seismic analyses do not directly probe key quantities such as the opacity
or the mean molecular weight. There will always be some degree of degeneracy in helioseismic
inferences, which leads to difficulties in pinpointing the exact causes of the discrepancies. Besides
helioseismic inversions, more direct constraints such as the so-called frequency ratios of low-order
pressure modes \cite{RoxburghRatios} to constrain the properties of the deep radiative layers of solar models. We illustrate in figure \ref{figRatios} the changes that can be observed by changing the opacity tables and chemical
abundances of solar models.

An example of such degeneracy is illustrated in Buldgen et al. \cite{Buldgen2019Sun}, where combinations of
various seismic inversions are used to draw conclusions on the solar modelling problem. Another
approach is to use seismic solar models to draw inferences on the properties of other “secondary”
thermodynamical quantities as in \cite{Shibahashi2006} for example. Other approaches include so-called extended
solar calibrations, as discussed in Section 4. These calibrations \cite{Ayukov2013, Ayukov2017}, using more constraints than
a classical standard model calibration, allow for the inclusion of “non-standard” corrections to the
models, such as modifications of the opacity or the nuclear reaction rates. Interestingly, they draw
similar conclusions to the study by Buldgen et al. \cite{Buldgen2019Sun} and other studies using seismic models or
ad-hoc corrections to the models. Perhaps one of the most significant conclusion from these studies is
that a revision of the opacity on a restricted domain in temperature will not be sufficient to reconcile
low metallicity models with helioseismology.

Combining the inferences from seismic solar models with various structural inversions to these
calibrations may perhaps provide a way to test with unprecedented thoroughness the ingredients of
solar models. For example, helioseismology has clearly proven to be able to probe the temperature and
chemical composition gradient just below the base of the convective envelope. Such investigations
could provide stringent constraints on the properties of overshooting in stellar envelopes and perhaps
a crucial reference point for more realistic prescriptions \cite{Xiong01, Rempel04, Zhang14, Gabriel2018}.

\section{Conclusion}
In this brief review, I have discussed some of the main inference techniques that have been used to
exploit the information of the solar global oscillations. More detailed discussions can be found in
more extended reviews (e.g. \cite{JCD2002Review, Thompson2003, Basu08, KosoReview}).

It is obvious that the main current issue in the analysis of the global solar oscillations is the inad-
equacy between the helioseismic constraints and standard solar models built using the revised solar
spectroscopic abundances. While it is certain that these determinations have their own uncertainties,
there seems to be no indication that further revisions would reincrease the solar metallicity back to
its value inferred from the empirical 1D model. Consequently, the low abundances are here to stay,
and it is up to seismologists and modellers to attempt to determine the origin of the discrepancies and
reduce them.

In this context, global helioseismology has a strong potential and a bright future. With a tremen-
dous amount of data to exploit, a large variety of new methods to implement and exploit, and a strong
implication of numerous groups involved in the refinements of the physical ingredients of the solar
and stellar models.

However, the stakes of the solar problem are not only a matter of detailed discussions about the
solar structure. Indirectly, a revision of the ingredients of solar models will snowball into a revision
of the model grids used widely in astrophysics to study all other stars in the Universe and is already
leading to intense discussions about the modelling of radiative opacities in stellar conditions. As such, the Sun remains to this day a wonderful laboratory of fundamental physics on both microscopic and
macroscopic scales.
\section*{Acknowledgements}
G.B. acknowledges support from the ERC Consolidator Grant funding scheme (project ASTER-
OCHRONOMETRY, G.A. n. 772293). This work is sponsored the Swiss National Science Foundation
(project number 200020-172505).
%
%
%



\end{document}